\documentclass[conference]{IEEEtran}

\ifCLASSINFOpdf
  \usepackage[pdftex]{graphicx}
  \graphicspath{{../pdf/}{../jpeg/}}
  \DeclareGraphicsExtensions{.pdf,.jpeg,.png}
\else
   \usepackage[dvips]{graphicx}
   \graphicspath{{../eps/}}
   \DeclareGraphicsExtensions{.eps}
\fi

\usepackage[cmex10]{amsmath}

\usepackage{algorithmic}

\usepackage{array}
\usepackage{mdwmath}
\usepackage{mdwtab}
\usepackage{eqparbox}

\usepackage{multirow}
\usepackage{xtab}

\begin{document}

\title{The Design and Implementation of an ANN-based Non-linearity Compensator of LVDT Sensor}
\author{  
  \IEEEauthorblockN{Prasant Misra$^\ddagger$ Santoshini Kumari Mohini$^{\dagger}$ Saroj Kumar Mishra$^{*\dagger}$}\\
  \IEEEauthorblockA{Robert Bosch Centre for Cyber Physical Systems, Indian Institute of Science, Bangalore, India$^\ddagger$}
  \IEEEauthorblockA{Accenture Services Pvt. Ltd., Bangalore, India $^{\dagger}$}
  \IEEEauthorblockA{Aricent Technologies, Chennai, India $^{*\dagger}$}  	
	Email: prasant.misra@rbccps.org$^\ddagger$, santoshini.mohini@gmail.com$^{\dagger}$, sarojkmishra@yahoo.com$^{*\dagger}$
}


\maketitle

\begin{abstract}
\boldmath
Linear variable differential transformer (LVDT) sensors are used in engineering applications due to their fine-grained measurements.
However, these sensors exhibit non-linear input-output characteristics, which decrease the reliability of the sensing system. 
The contribution of this article is three-fold. 
First, it provides an experimental study of the non-linearity problem of the LVDT.
Second, it proposes the design of a functional link artificial neural network (FLANN) based non-linearity compensator model for overcoming it.
Finally, it validates the feasibility of the solution in simulation, and presents a proof-of-concept hardware implementation on a SPARTAN-II (PQ208) FPGA using VHDL in Xilinx.
The model has been mathematically derived, and its simulation study has been presented that achieves nearly $100 \%$ linearity range.
The result obtained from the FPGA implementation is in good agreement with the simulation result, which establishes its actualization as part of a general manufacturing process for linearity compensated LVDT sensors.
\end{abstract}

\section{Introduction}

Linear variable differential transformer (LVDT) sensors are utilized in various control system applications for measuring displacement, pressure, force, and other physical quantities.
These sensors provide numerous advantages in the form of fine-grained resolution and precise measurements, friction-free operating that increases its operational span, fast response, high sensitivity, and robust operation under wide temperature ranges and environmental conditions \cite{Neubert2003}.
However, an inherent problem is that they exhibit non-linear input-output characteristics, which lead to erroneous displacement recordings, thereby decreasing the reliability of the sensing system.  
Conventionally, obtaining high linearity working range during their fabrication in the factory requires sophisticated machinery.
Moreover, it is quite difficult to achieve fine tuning of every sensor manufactured to exhibit equal linear properties. 
Hence, the users have to undertake the tedious job of pitch calibration by adjusting the screw gauge on this device. 
Even after manual calibration, these devices may exhibit non-linear behaviour due to inherent difference in their characteristics, variation in environmental conditions, aging; or simply due to human errors.
This results in the decrease of the usable operating range of the device, and also affects the system accuracy.
Hence, there is a need for an automated process to calibrate each LVDT sensor.
\newline
\indent
In this paper, we provide a detailed insights into this problem, and make the following contributions:
\begin{enumerate}
	\item We present a study of the non-linearity problem of the LVDT by gathering experimental data from an off-the-self sensor, and demonstrate its limited operational range due to its non-linear input-output characteristics.
	\item We propose an artificial neural network (ANN) based inverse modeling approach for overcoming this problem. 
	We utilize a variant of the traditional ANN called functional link artificial neural network (FLANN). 
	This model has been explained through mathematical derivation, and has been verified through simulation in MATLAB, which achieve nearly 100\% linearity.
	\item We propose an algorithm for the proof-of-concept hardware implementation of this scheme on a SPARTAN-II (PQ208) field programmable gate array (FPGA) using VHDL in Xilinx.
\end{enumerate}
In addition, the the lessons and experiences may be helpful to other engineers who are working on similar problems and projects.
\newline
\indent
The remainder of the article is arranged as follows.
Sections \ref{sec:overview} and \ref{sec:pre_exp} present a general study of the LVDT sensor.
Section \ref{sec:system_model} and \ref{sec:FPGA_impl} explores the mathematical design of an inverse modeling approach for automatic calibration using FLANN.
Section \ref{sec:evaluation} presents the evaluation results of both the simulation studies and FPGA implementation.
Section \ref{sec:related_work} follows it up by providing a concise background and overview of related work.
The final Section \ref{sec:conclusion} suggests various possible improvements to the design, and concludes with a summary of the areas covered in the paper.

\begin{figure*}[t]
\begin{center}
\begin{tabular}{cc}
\includegraphics[width=3in]{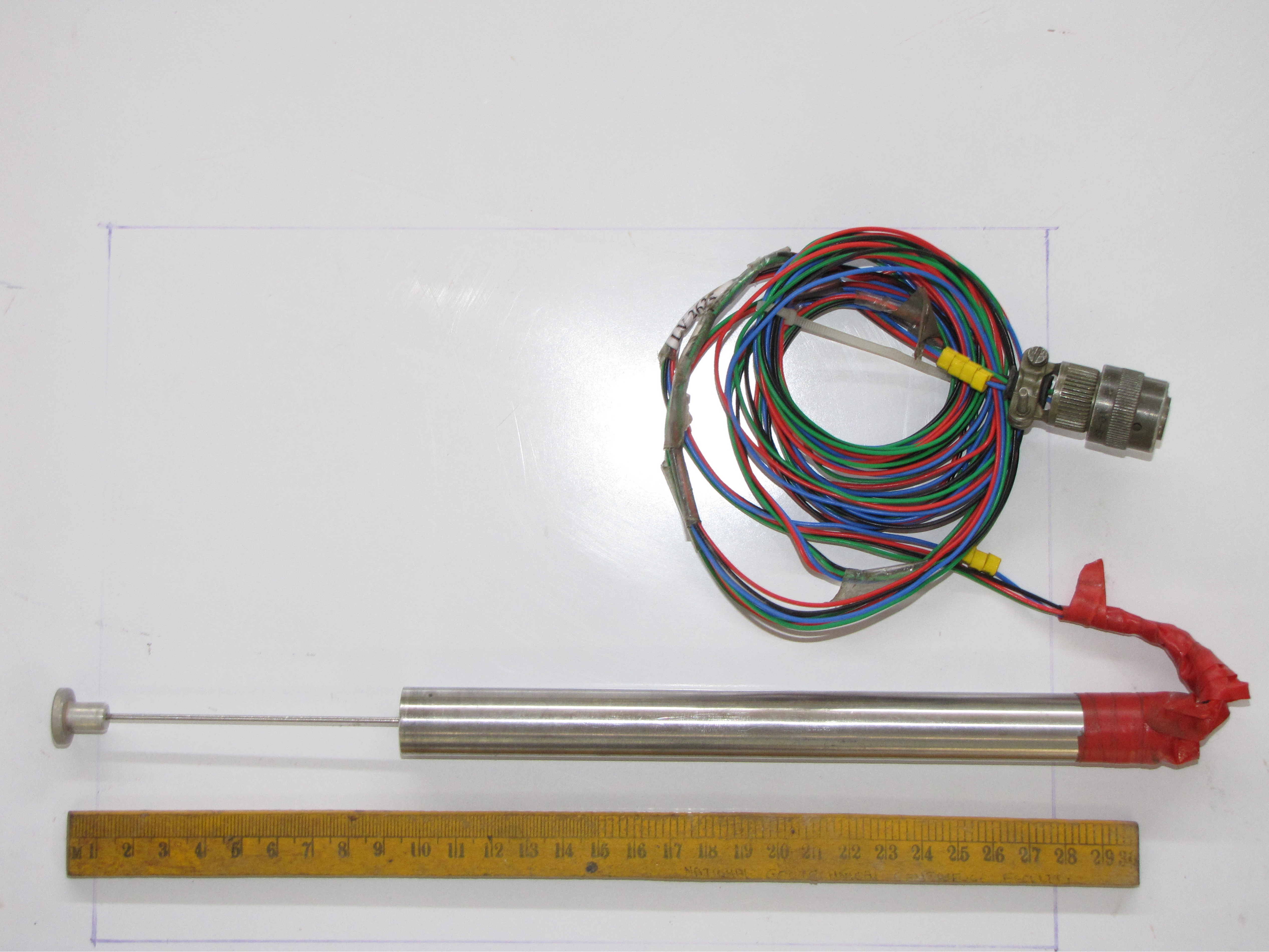} & \hspace{1cm} \includegraphics[width=3in]{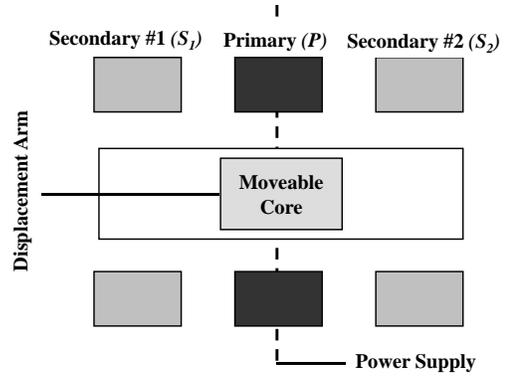} \\
(a) & \hspace{1cm} (b) \\
  \end{tabular}
 \end{center}
 \caption{(a): A Linear Variable Differential Transformer. (b): Horizontal cross-sectional view of a LVDT. }
\label{fig:LVDT_overview}
\end{figure*}
 
\section{LVDT Overview} \label{sec:overview}

A LVDT is a displacement sensor that measures physical movement (or displacement), and represents this change as an output voltage.
It consists of three coils: a single primary coil $P$ (known as the emitter coil), and two secondary coils $S_{1}$ and $S_{2}$ (known as receiver coils) wound on a cylindrical former (Figure \ref{fig:LVDT_overview}-(b)). 
The two secondary coils are identical (i.e. they have equal number of turns), counter-wound (i.e. if coil windings of $S_{1}$ are clockwise, then the windings on $S_{2}$ are counter-clockwise, or vice-versa ), and are placed on either side of the primary winding. 
The primary winding is connected to a power source (either alternating or direct). 
A movable soft iron core is placed inside the former, and a displacement arm is attached it. 
The movement of the arm displaces the primary coil, which induces a signal on the two secondary coils.
This signal has an amplitude that is nearly proportional to the displacement of the core. 
A reverse movement of the arm (and subsequently the core) would result in the change of the sign of the signal.
This measure of the displacement is converted into its respective voltage output, and is recorded by the data acquisition unit (DAQ). 
Typically, the DAQ is attached to a computer that has the specified interface to mount the DAQ card. 
The analog-to-digital converter (ADC) unit of the DAQ is responsible for digitizing the received analog signals, and passing it onto the display software.
\newline
\indent
The LVDT shown in Figure \ref{fig:LVDT_overview}-(a) was used in our experiments. 
It did not have any external power supply. 
The cable leading from the LVDT was used both for powering the sensor, as well as for data transfer. In this manner, it can be viewed as a USB drive connected to a specific port on the CPU. 
The following section describes the working principle of this sensor.

\subsection{Working Principle} \label{sec:working}

A voltage of $V_{in}$ is applied across the primary coil.
Depending on the position of the core (whose movement is caused due to the displacement in the actuator arm) with respect to the primary coil, inductance takes place on both the secondary coils generating voltages $V_{s1}$ and $V_{s2}$. 
The final voltage output $V_{out}$ from the LVDT is the difference of the voltages attained by the the secondary coils, i.e. $V_{out}=V_{s1} - V_{s2}$. $V_{out}$ is a direct representation of the displacement of the actuator arm.
\newline 
\indent
The operation of the LVDT can be best described by considering the following 3 distinct positions of the core: 
\vspace{0.5mm}
\newline
\noindent
\textit{Stage 1: Core is at its normal (NULL) position.} The flux linking with both the secondary windings is equal ($V_{s1}=V_{s2}$), and hence, equal emfs are induced in them. 
Thus, $V_{out}=0$ at the NULL position, and so is the displacement. 
\vspace{0.5mm}
\newline 
\noindent
\textit{Stage 2: Core is moved to the left of the NULL position.} In this case, more flux links with winding $S_{1}$ and less with windings $S_{2}$. Accordingly, the output voltage of the secondary winding $S_{1}$ is more than $S_{2}$ ($V_{s1} > V_{s2}$). 
Thus, the output voltage is in phase with the primary voltage. 
The displacements recorded along this direction would all be of the same sign, with an increase in value, as the core is displaced further away from the NULL position.
\vspace{0.5mm}
\newline 
\noindent
\textit{Stage 3: Core is moved to the right of the NULL position.} In this case, the flux linking with winding $S_{2}$ becomes larger than that linking with $S_{1}$. Accordingly, the output voltage of the secondary winding $S_{2}$ is greater than $S_{1}$ ($V_{s2} > V_{s1}$). Thus, the output voltage is $180^{o}$ out of phase with the primary voltage. Consequently, the displacements recorded along this direction would all be of the same sign, but opposite to the sign of the values recorded in Case 2.

\section{Preliminary Experiments} \label{sec:pre_exp}

In this section, we provide an overview of the experimental work performed for the empirical data collection from a LVDT, and our own experiences in working with it.
The traces, in the form of input-output characteristics, have been analyzed and discussed, thereby highlighting the issues that need to be addressed when modeling the non-linearity compensator unit for this sensor.

\subsection{Experimental Setup and Results} \label{sec:pre_exp1}

In order to create an accurate non-linearity compensator model, we required a comprehensive database of input-output characteristic traces of a LVDT.
For this task, we collected data from a simple LVDT having the following specifications:
\\Internal diameter of the core: $4.4$ mm.
\\External diameter of the core: $5.0$ mm.
\\Core length: $60.0$ mm.
\\Number of winding turns on primary coil: $1500$.
\\Number of winding turns on each secondary coil: $3300$.
\\The two secondary coils are separated by a Teflon ring.
\\Excitation frequency: $5.0$ kHz.
\\Excitation voltage (peak-to-peak): $10.0$ $V_{pp}$.
\newline
\indent
The experimental setup consisted of three units: desktop-computer-based controller, stepper-motor-based ($7.5^{o}$/step rotation) displacement actuator, and an off-the-shelf LVDT sensor.
The stepper-motor performs a regulated displacement of the core of the LVDT.
It is under the control of a computer program, and provides a linear displacement of $1.0$ mm to the LVDT core, at each stepper rotation.
This actuator system has been calibrated to produce a linear response.
It has been programmed to perform displacement, both in the forward and reverse direction, in order to get a trace of the symmetrical readings (both positive and negative) about the NULL position. 
The differential output voltage $v$ of the LVDT is recorded for every displacement of $x$.
The experimentally collected data is presented in Table \ref{tab:exp_measured_data}.

\begin{table}[t]
  \scriptsize
	\centering	
	\caption{Experimental Measured Data}
	\label{tab:exp_measured_data}
		\begin{tabular}{|p{2.5cm}|p{2.5cm}|}
			\hline
		\textbf{Displacement (mm)} & \textbf{Demodulated voltage output (V)} 
		\\ \hline	\hline
	  -30 & -5.185 \\ \hline
	  -25 & -5.017 \\ \hline
	  -20 & -4.717 \\ \hline
	  -15 & -4.039 \\ \hline
	  -10 & -2.896 \\ \hline
	  -5  & -1.494 \\ \hline
	  0 (Null position) & 0.001 \\ \hline
	  5   & 1.462 \\ \hline
	  10  & 1.810 \\ \hline
	  15  & 3.962 \\ \hline
	  20  & 4.799 \\ \hline
	  25  & 5.225 \\ \hline
	  30  & 5.276 \\ \hline 	  
	  \end{tabular}
\end{table} 

\subsection{Analysis and Discussion}

Let us define \textit{linearity} with respect to a LVDT sensor. 
As we have explained in Section \ref{sec:working}, a NULL (or zero) point is a position in the displacement of the core, where both the displacement, and its corresponding output voltage is zero.
Hence by \textit{linearity}, we means: \textit{irrespective of the displacement of the actuator arm to the left or right of this NULL point, the  output voltage recorded by the LVDT in response to this movement, should be same in magnitude.}
\newline
\indent
An analysis of the experimental data trace shows two important observations. 
First, the NULL point does not record a perfect zero voltage output, and registers a value of 0.001V at this dormant position. 
Second, for the same amount of displacement on either side of the core, the scalar magnitude of the voltage output from the LVDT are not the same. 
For example, a displacement of $10$ mm in both the directions (forward and reverse) from the core, shows values of $2.896$ V and $1.810$ V respectively (Table \ref{tab:exp_measured_data}). 
However, according to the working principle of LVDT, they both should have been the same (in magnitude), though with a different sign to represent the direction of motion. 
\textit{This deviation in the input-output characteristics of the LVDT is the problem that we seek to model.}
\newline
\indent
The following points summarize our findings from the experimental study.
\begin{itemize}
	\item The amount of voltage change is proportional to the amount of movement of the core, and therefore is an indication for it.      
	\item The direction of motion is inferred from the increase/decrease of the output voltage.
	\item The output voltage is a linear function of the core displacement within a limited range of motion. 
				Beyond this range of displacement, the characteristic curve starts to deviate from the straight line.
	\item There exists a small voltage at the NULL position, which ideally should be zero.
\end{itemize}

\begin{figure*}[t]
\begin{center}
\begin{tabular}{cc}
\includegraphics[width=3.25in]{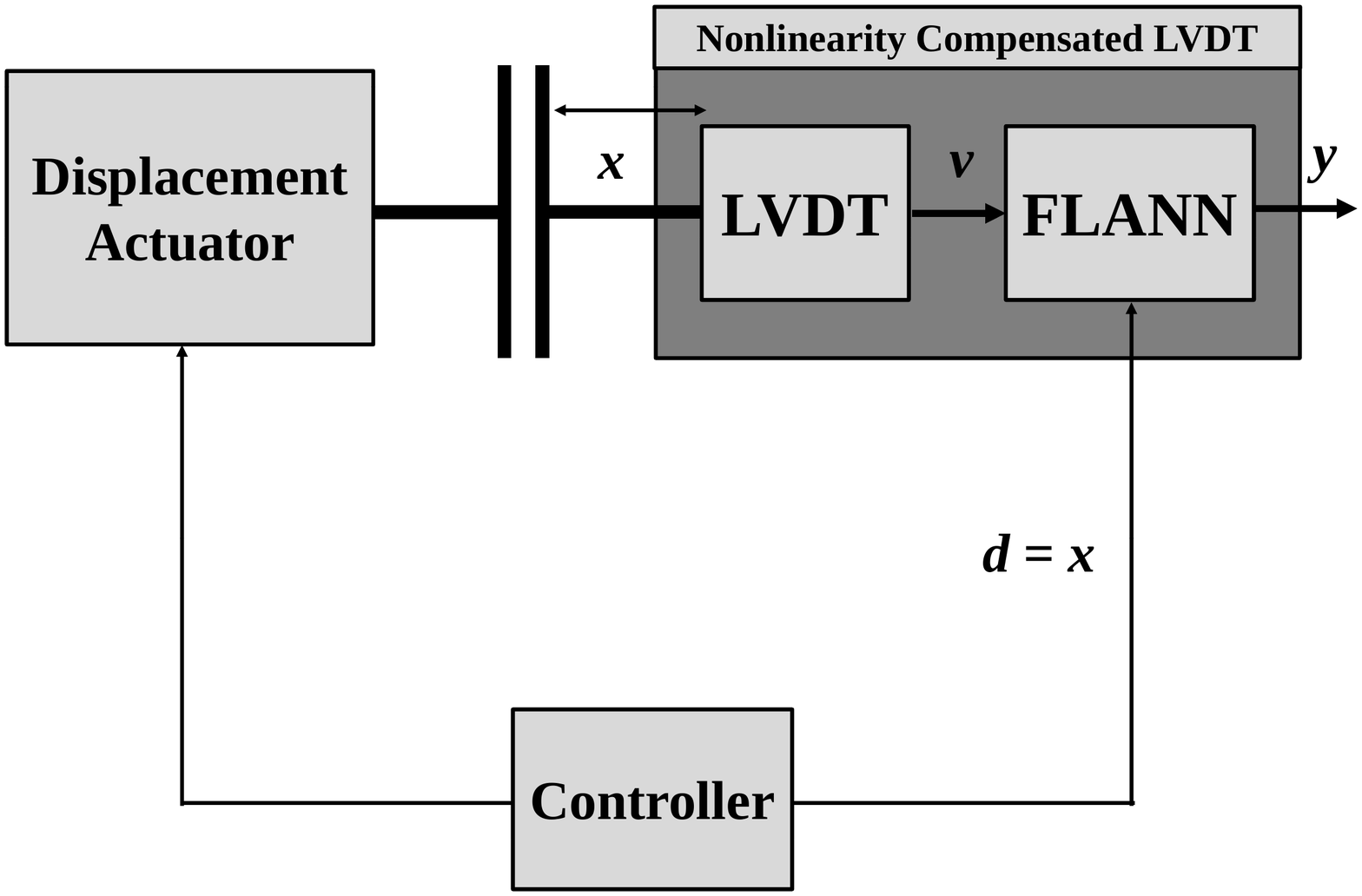} & \includegraphics[width=3.25in]{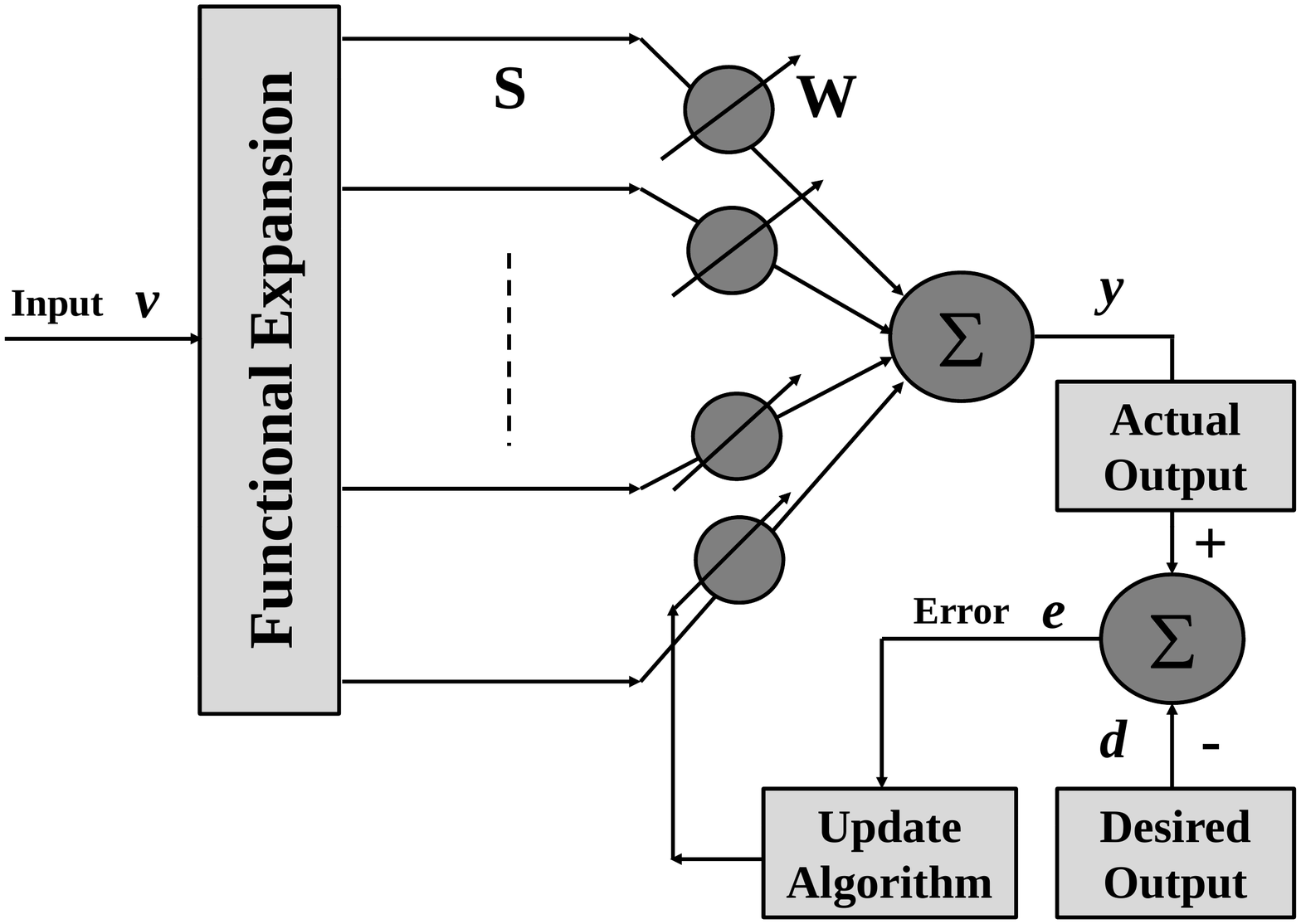} \\
(a) & (b) \\
  \end{tabular}
\end{center}
 \caption{(a): Scheme of a non-linearity compensator of LVDT. (b): Structure of the FLANN model.}
\label{fig:experimental_setup}
\end{figure*}

\section{System Model} \label{sec:system_model}
 
We utilize a simple design of \textit{inverse modeling} to compensate for the non-linearity exhibited by the LVDT. 
The basic functionality of an inverse model is to generate a mirror replica of the system under consideration. 
The non-linear LVDT sensor is cascaded with the adaptive inverse model to achieve overall linearity.
\newline
\indent
The proposed model of this experimental setup has been shown in Figure \ref{fig:experimental_setup}-(a).
It consists of all the units described in Section \ref{sec:pre_exp1}.
The only addition to it shall be an adjunct electronic element that replicates the functionality of the inverse model.
The displacement actuator displaces the core of the LVDT by a distance of $x$. 
It is under the control of a main controller that provides actuating signals for the controlled displacement of the core of the LVDT.
The corresponding nonlinear output $v$ is provided as input to this electronic element, which generates the output $y$, which resembles the displacement $x$ recorded by the controller.
\newline
\indent 
We achieve this functionality of the inverse model through the realization of an artificial neural network (ANN), which dynamically learns the system, and generates its inverse characteristics. 
The ANN utilized in our model is a Functional Link ANN (FLANN) \cite{Patra1999}. 
The prime advantages of the FLANN are: less computational complexity, ease of implementation and higher linearity range, compared to the multilayer perceptron (MLP) \cite{JCPatra1999} and radial basis function (RBF) based ANNs. 
Additionally, we wanted it to be simple in design, so that in our future implementation, it could be easily designed, programmed and configured on an FPGA chip.

\subsection{FLANN}

Figure \ref{fig:experimental_setup}-(b) shows the structure of a FLANN.
It is a single layer ANN with no hidden layers.
The voltage at the output of the LVDT ($v$), which is nonlinear, is provided as input to the FLANN model.
It is subject to functional expansions using mathematical series (such as: trigonometric, power series, tensor, outer product).
These functional links acts as a pattern of linearly independent functions, and these functions are evaluated with this pattern as an argument. 	
In our implementation, we utilize trigonometric expansions, because they provide better nonlinearity compensation as compared to other mathematical series \cite{Patra1999}. 
These expansions are multiplied with a set of neural weights, and finally added to produce the output of the inverse model.
It is compared with the desired signal (actuating signal of the displacement actuator) to derive the error signal. 
These weights are updated in order to minimize the mean square error (MSE) \cite{Sterns1985}. 
The process of training is repeated until the MSE reaches a minimum threshold level, beyond which it does not improve the estimates.
The dual combination of the LVDT and the FLANN represents a linear sensor with increased linearity range. 

\subsection{Derivation of the model} \label{sec:derivation}

The general learning technique of an ANN consists of interpolating a continuous, multivariate function $f(x)$ through an approximating function $f_{approx}(x)$. 
In a FLANN, $f_{approx}(x)$ is represented using a set of basis functions $\phi$, and a fixed number of weight parameters $W$. 
$\phi$ is choice based, which limits the learning problem to finding $W$, that provides the best approximation of $f(x)$ for a set of input-output.
\newline
\indent
Let the $N$ input elements $\{v_{1}, v_{2}, v_3,...,v_{N}\}$ to the FLANN be represented as matrix \textbf{V} of size $N\times 1$. 
Thus, the $n^{th}$ element can be given as $v_{n}$, where $1\leq n \leq N$. 
Every element $v_{n}$ is expanded to form $M$ elements such that the resultant matrix \textbf{S} has dimensions $N\times M$. 
This nonlinear expansion is achieved through the set of basis functions $B=\{\phi_{i}\}$ with the following properties: 
\begin{enumerate}
	\item $\phi_{1}=1$.
	\item If $\sum_{i=1}^N w_{i}\phi_{i}=0$, then $w_{i}=0$ for all $i=\{1,2,3,...,j\}$.
\end{enumerate}
\vspace{2mm}
The FLANN consists of $N$ basis functions: $\{\phi_{1},\phi_{2},\phi_{3},...,\phi_{N}\}\in B$. 
\newline
\indent
Each element $v_{n}$ in \textbf{V} undergoes trigonometric expansion using the following equation.
\begin{equation}
S_{i}=\left\{ \begin {array}{ll} v_{i} & \mbox{$i = 1$} \\
\sin(m\pi v_{i}) & \mbox {$i > 1, i = even (2,4,...,M)$}\\
\cos(m\pi v_{i}) & \mbox {$i > 1, i = odd (1,3,...,M+1)$}
\end {array} \right.
\end{equation}
where $m=1,2,...,M/2$. In this implementation, we have chosen $M$ to be an even number. 
Thus, $[S_{i}]_{i=1}^{M+2}$ can be represented as a matrix \textbf{S} of size $N \times P$ where $P=M+2$.
\newline 
\indent
Let $w=[w_{1},w_{2},w_{3},...,w_{P}]$ be the weight vector of this FLANN having $P$ elements. 
Here, we use a heuristic approach to assign values to $w$. 
Rather than choosing any random value for $w$, we assign a value of unity (1) to every element of the the weight vector. 
This is done to get the highest level (or values) of the functional expansion, which would facilitate in achieving the estimated weight values in less iterations.
\newline 
\noindent
The output at each iteration $k$ is given as:
\begin{equation}
	y(k) = \sum_{p=1}^{P}s_{p}.w_{p}
\end{equation}
In matrix notation, it is represented as:
\begin{equation}
	Y = S_{P}.W_{P}^{T}
\end{equation}
The corresponding error between the estimated and the desired output is given by:
\begin{equation}
	\epsilon(k)=d(k)-y(k)
\end{equation}
where $d(k)$ is the desired signal, which is the same as the control signal given to the displacement actuator.\\ 
Let the cost function or the residual noise power at the $k_{th}$ iteration be denoted as $\xi(k)$ and is given by:
\begin{equation}
	\xi(k) = \sum_{j \in P}\epsilon_{j}^{2}(k)
\end{equation}
The weight vector $\textbf{W}$ of the FLANN is updated using the least-mean-square (LMS) algorithm, in order to minimize the mean square error $\xi$ and is given by:
\begin{equation}\label{eq:1}
	w(k+1) = w(k) + \Delta w(k) 
\end{equation}

$\Delta w(k)$ is given by:
\begin{equation}
\begin{array}{lll}
\label{eq:}
  \Delta w(k) & = & - \frac{\eta}{2}.\frac{\partial \xi(k)}{\partial w(k)} \\
              & = & \eta \epsilon(k) s(k)
\end{array}
\end{equation}
\noindent
where $\eta$ is the learning rate parameter ($0 \leq \eta \leq 1$), or the step-size that controls the convergence speed of the LMS algorithm, and $\nabla$  is the instantaneous estimate of the gradient of $\xi$ with respect to $w(k)$.
Thus, equation \ref{eq:1} can be given as:
\begin{equation}
	w(k+1) = w(k) + \eta \epsilon(k) s(k)
\end{equation}
The weight vector (or the respective neural weights) can be obtained as:
\begin{equation}
  W={S}^{-1}.Y
\end{equation}

\section{FPGA Implementation} \label{sec:FPGA_impl}

This section presents our proposed algorithm for the proof-of-concept hardware implementation of the ANN-based non-linearity compensator design.
We do not claim it to be optimal or the best.
Nevertheless, it establishes the practical implementability of our solution, and provides a good learning experience.
\newline
\indent
Our algorithm requires three functional blocks for its implementation.
\begin{itemize}
	\item Expansion block: Performs a functional expansion of the input values. 
	      Its working is similar to a de-multiplexer where a single input would be converted into many outputs.
	\item Multiplication block: Performs the multiplication of the neural weights with the functionally expanded signals, and is based on the algorithm discussed in \cite{Santorothesis1989}.
	\item Addition block: Performs the final addition of the various intermediate signals, and is based on the algorithm presented in \cite{Meher2004}.
\end{itemize}
We use 18-bit floating point numbers to accurately represent every decimal number in its equivalent binary form (preserving its decimal point). 

\subsection{Proposed Algorithm of the Inverse Model}

\begin{figure}[h]
\begin{center}
\begin{tabular}{c}
\includegraphics[width=3.5in]{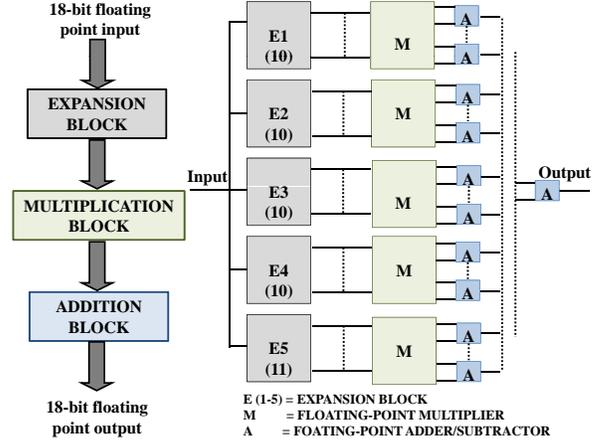} \\
\end{tabular}
\end{center}
\caption{FPGA implementation. (Left): Design flow. (Right): Internal architecture.}
\label{fig:FPGA_imp}
\end{figure}

The algorithm has been implemented with the help of a Look-up table.
It is designed for a few specific values only.
These values were taken from the experimentally measured data set specified in Table \ref{tab:exp_measured_data}. 
They were converted to their respective 18-bit floating point format, through manual calculation, and then stored in the table.
When these values are provided as input to the setup, it searches this table for its corresponding floating point representation, and then proceeds to the next stage.
Similarly, the final output, which is in 18-bit floating point format, is evaluated for its decimal equivalent value.
The design flow has been divided into three stages as shown in Figure \ref{fig:FPGA_imp}.

The total number of expansions originating from the FLANN is 51.
The motivation for using 51 functional expansion shall be discussed in Section \ref{sec:sim}.
The description of the different stages are as follows:

\subsubsection{Stage 1 : Expansion}

The objective of this stage is to perform a functional expansion of the input into 51 signals. 
The output from the LVDT is fed as a 18-bit floating point input to the expansion block. 
\newline
\indent
It consists of five expansion sub-blocks. 
Out of the five sub-blocks, four of these take a single input, and produce 10 outputs, while the fifth block produces 11 outputs. 
The total number of expansions originating from the FLANN block is 51 [10 outputs x 4 sub-blocks + 11 outputs x 1 sub-block].
Each output is in 18-bit floating point format.
\newline 
\indent
Expansion sub-block E1 produces outputs based on the following solution set.\\
$\{v, \sin(\pi v), \cos(\pi v), \sin(2\pi v), \cos(2\pi v), \sin(3\pi v), \cos(3\pi v), \\ \sin(4\pi v), \cos(4\pi v), \sin(5\pi v)\}$
\newline 
\indent
Expansion sub-block E2 produces outputs based on the following solution set.\\
$\{\cos(5\pi v), \sin(6\pi v), \cos(6\pi v), \sin(7\pi v), \cos(7\pi v), \sin(8\pi v), \\ \cos(8\pi v), \sin(9\pi v), \cos(9\pi v), \sin(10\pi v)\}$
\newline 
\indent
Expansion sub-block E3 produces outputs based on the following solution set.\\
$\{\cos(10\pi v), \sin(11\pi v), \cos(11\pi v), \sin(12\pi v), \cos(12\pi v), \\ \sin(13\pi v), \cos(13\pi v), \sin(14\pi v), \cos(14\pi v), \sin(15\pi v)\}$
\newline 
\indent
Expansion sub-block E4 produces outputs based on the following solution set.\\
$\{\cos(15\pi v), \sin(16\pi v), \cos(16\pi v), \sin(17\pi v), \cos(17\pi v), \\ \sin(18\pi v), \cos(18\pi v), \sin(19\pi v), \cos(19\pi v), \sin(20\pi v)\}$
\newline 
\indent
Expansion sub-block E5 produces outputs based on the following solution set.\\
$\{\cos(20\pi v), \sin(21\pi v), \cos(21\pi v), \sin(22\pi v), \cos(22\pi v), \\ \sin(23\pi v), \cos(23\pi v), \sin(24\pi v), \cos(24\pi v), \sin(25\pi v)\}$

\begin{figure*}[t]
\begin{center}
\begin{tabular}{ccc}
\includegraphics[width=2in]{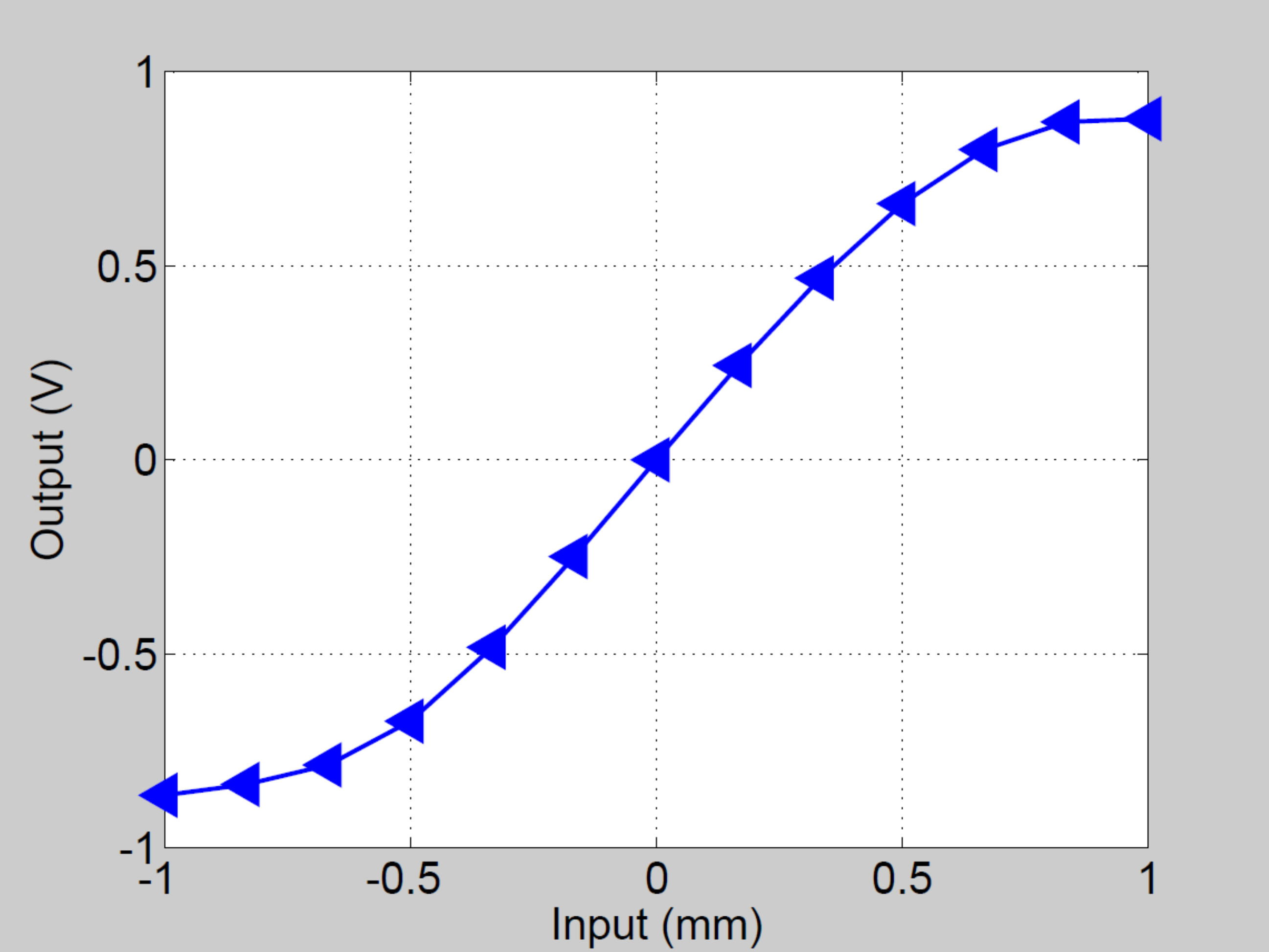} & \includegraphics[width=2in]{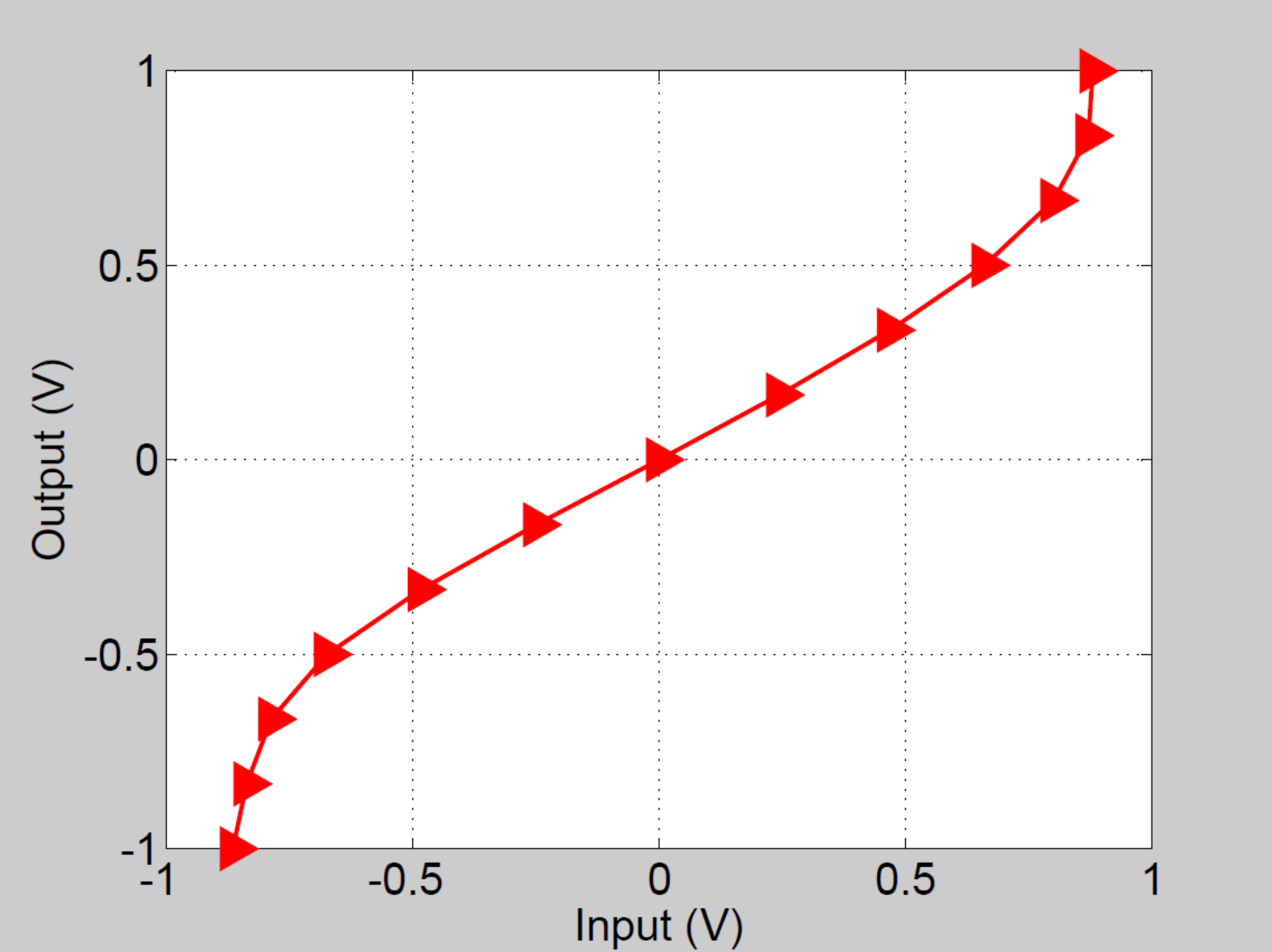} 
& \includegraphics[width=2in]{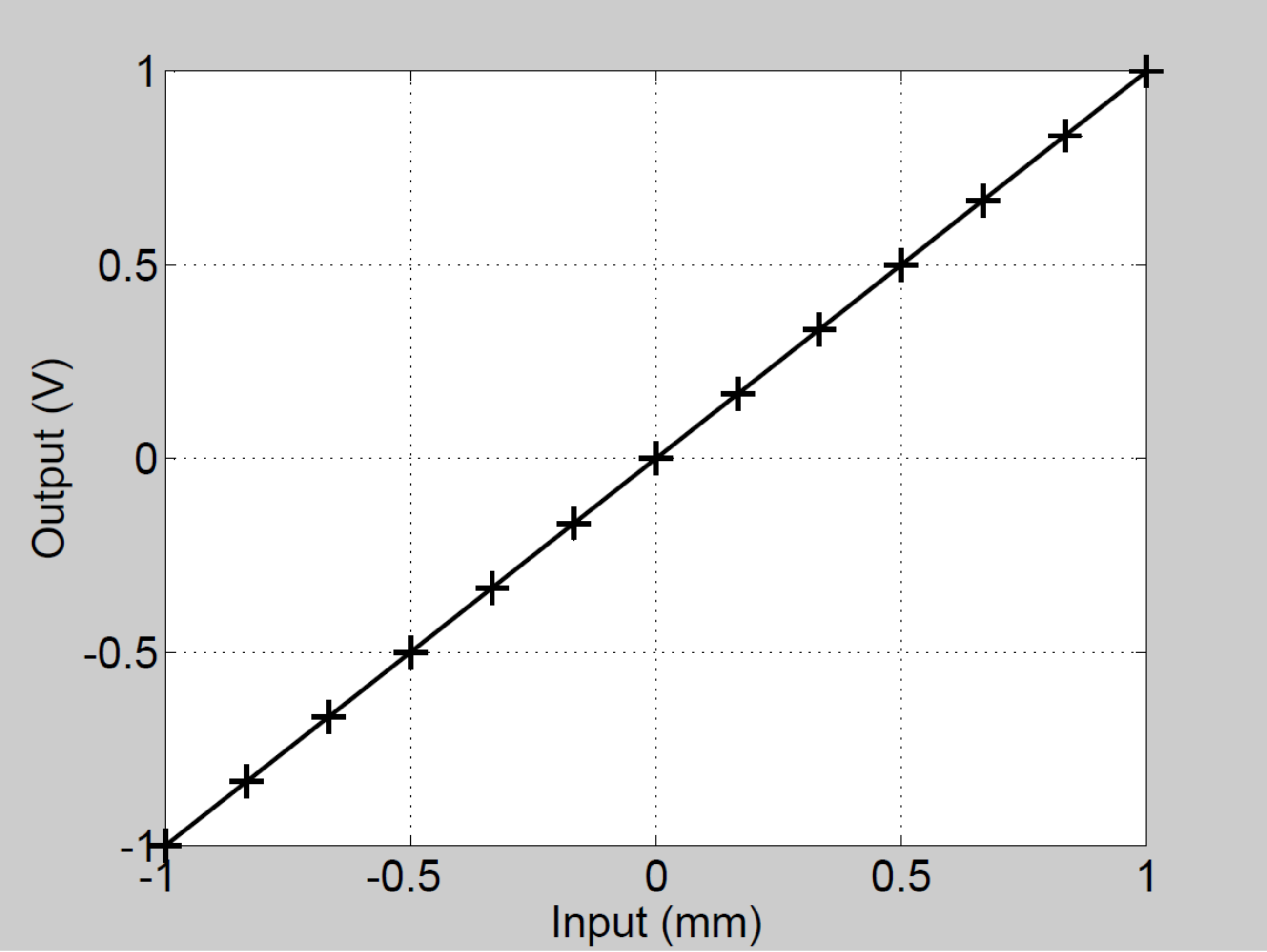} \\
(a): LVDT (nonlinear). & (b): FLANN (Inverse Model). & (c): Overall system response: Linear. \\\\
\includegraphics[width=2in]{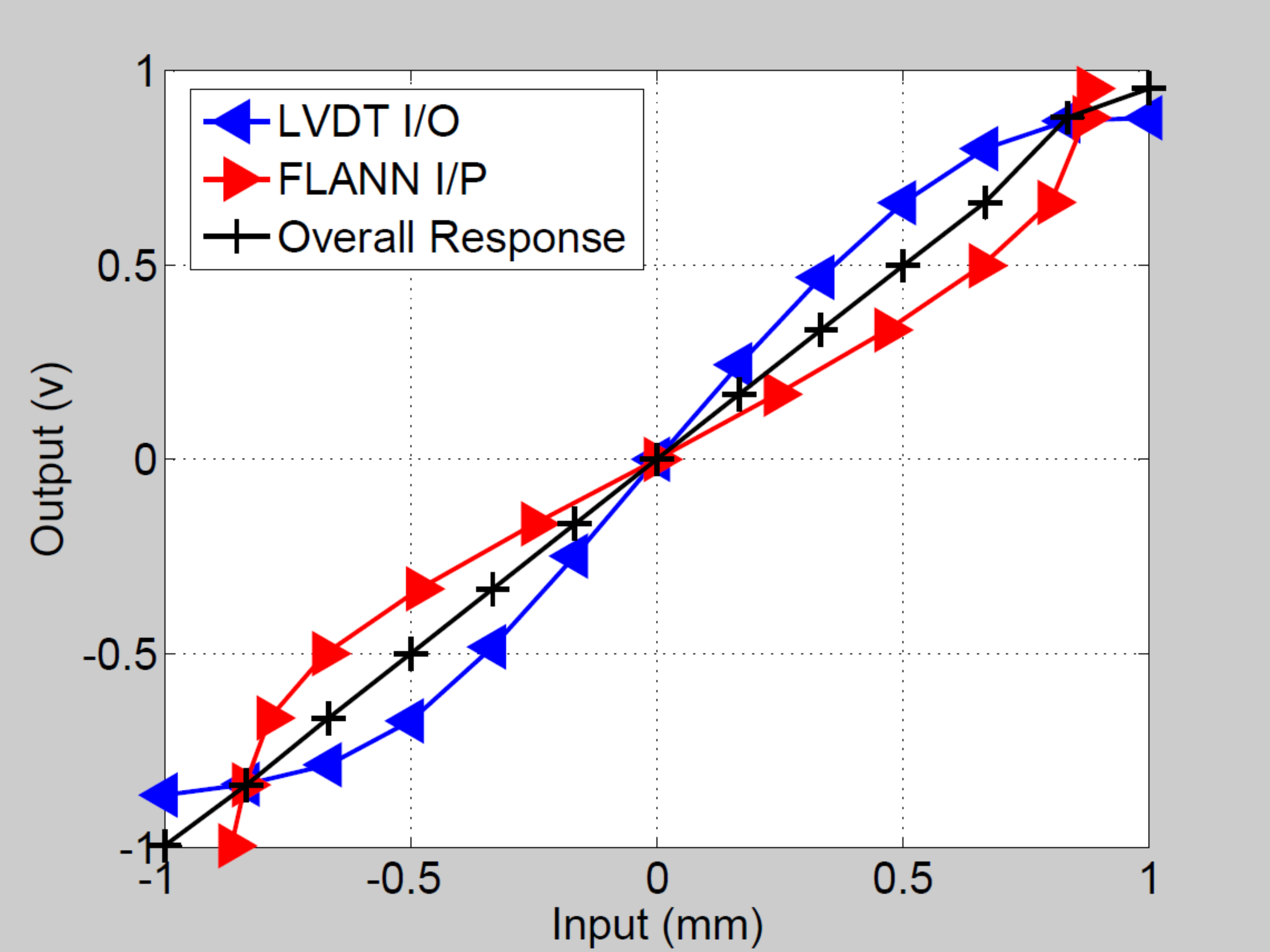} & \includegraphics[width=2in]{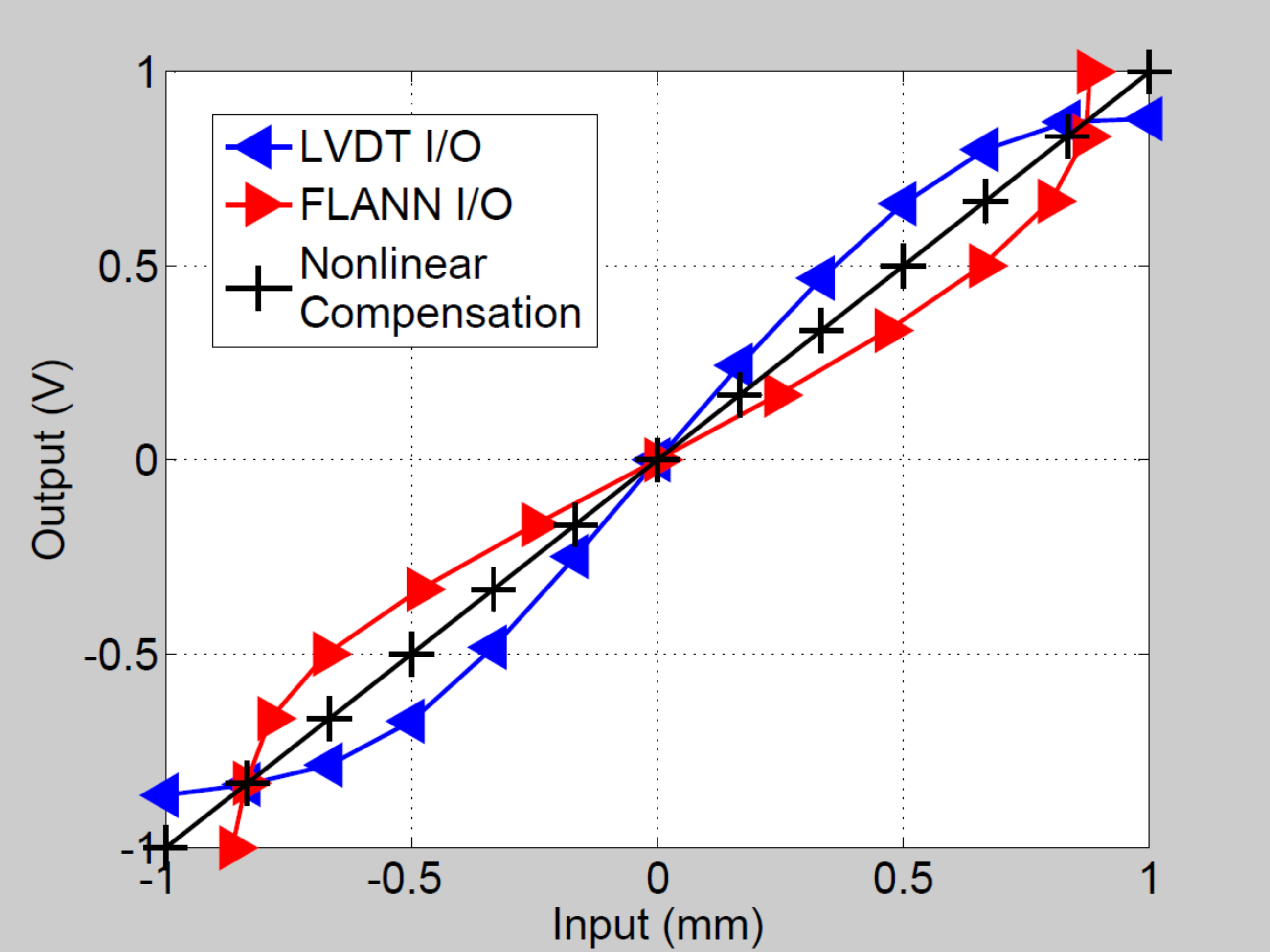} 
& \includegraphics[width=2in]{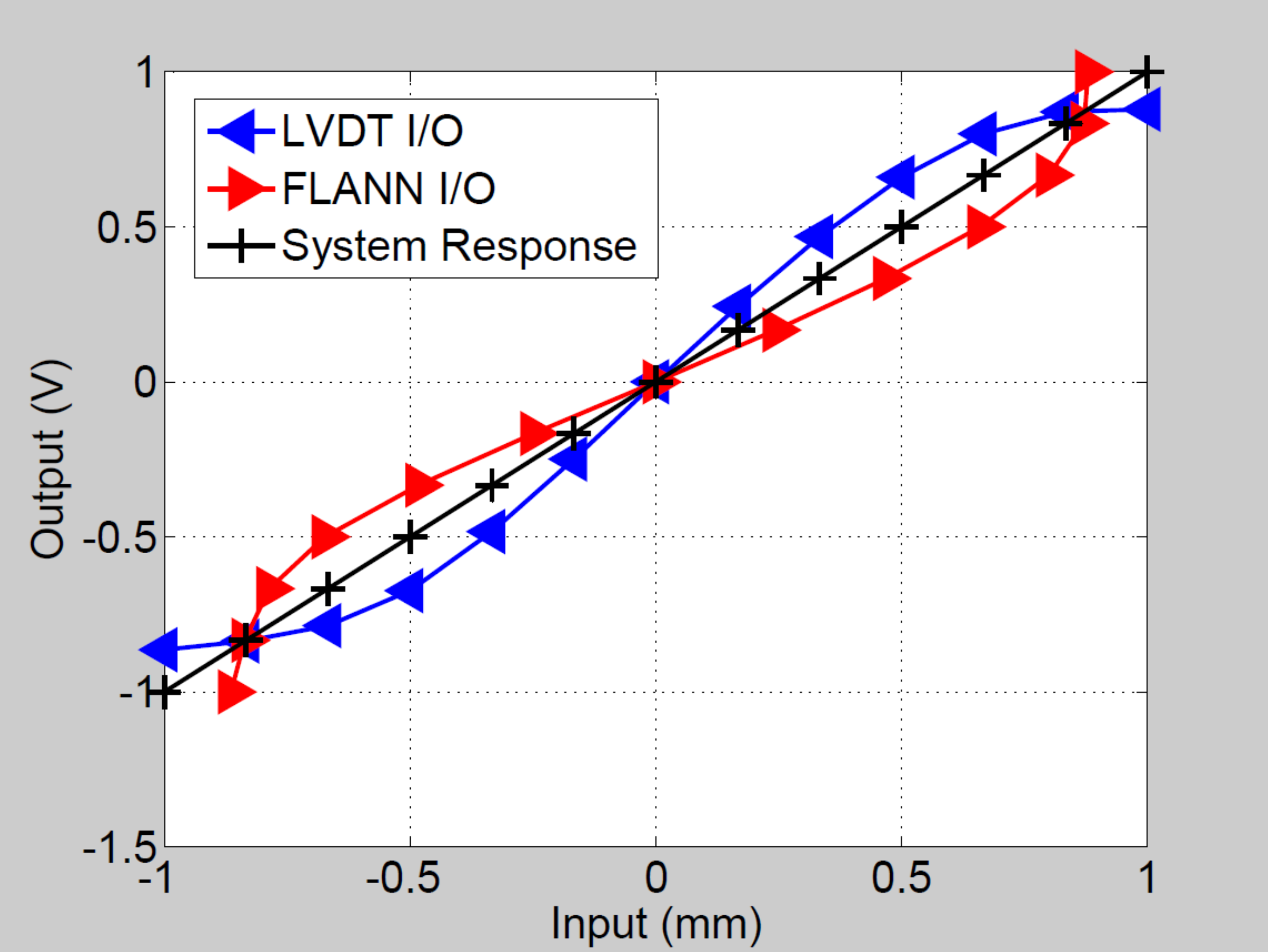} \\ 
(d): FE: 25. & (e): FE: 51. & (f): FE: 61. \\\\
\includegraphics[width=2in]{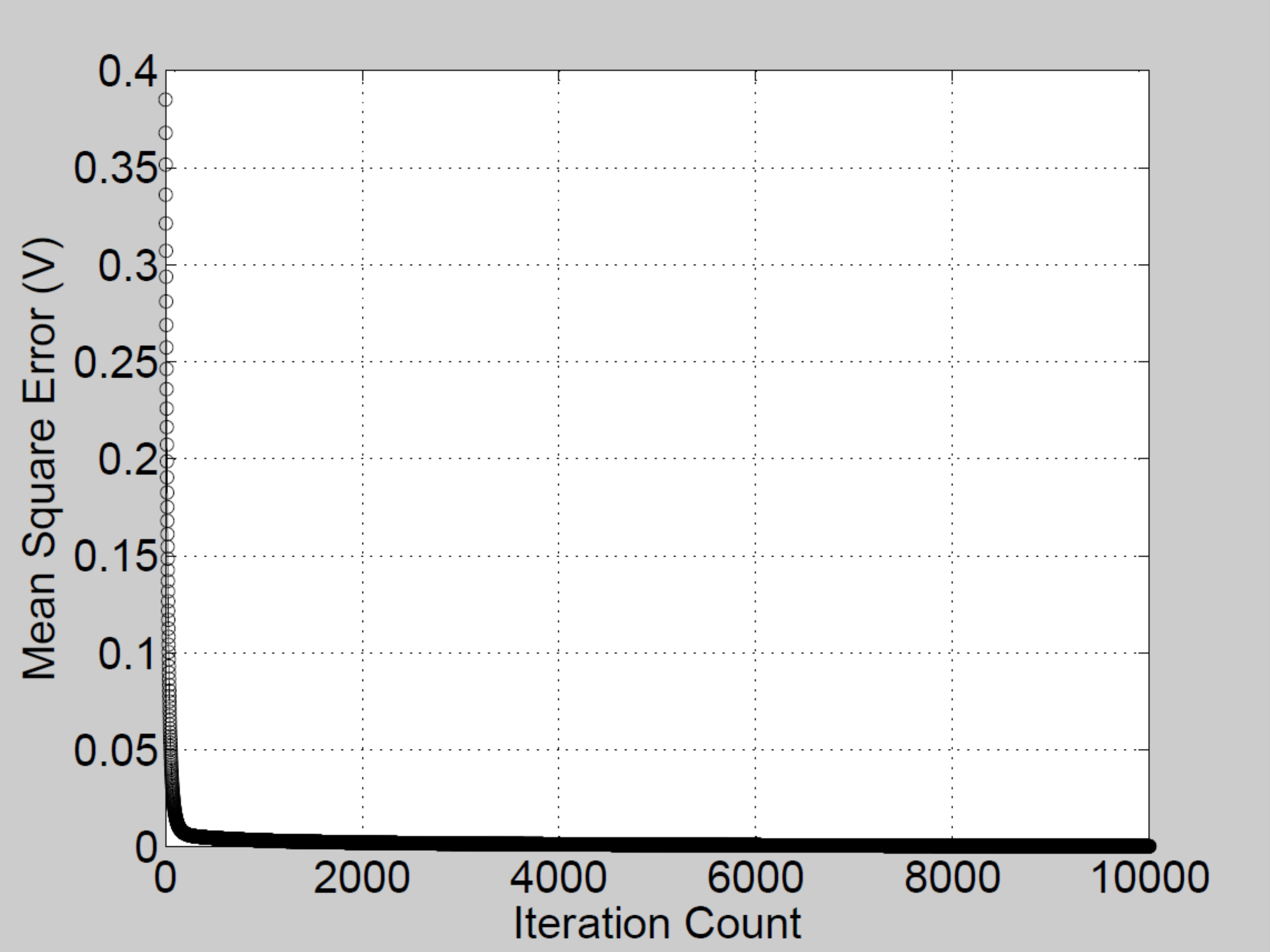} & \includegraphics[width=2in]{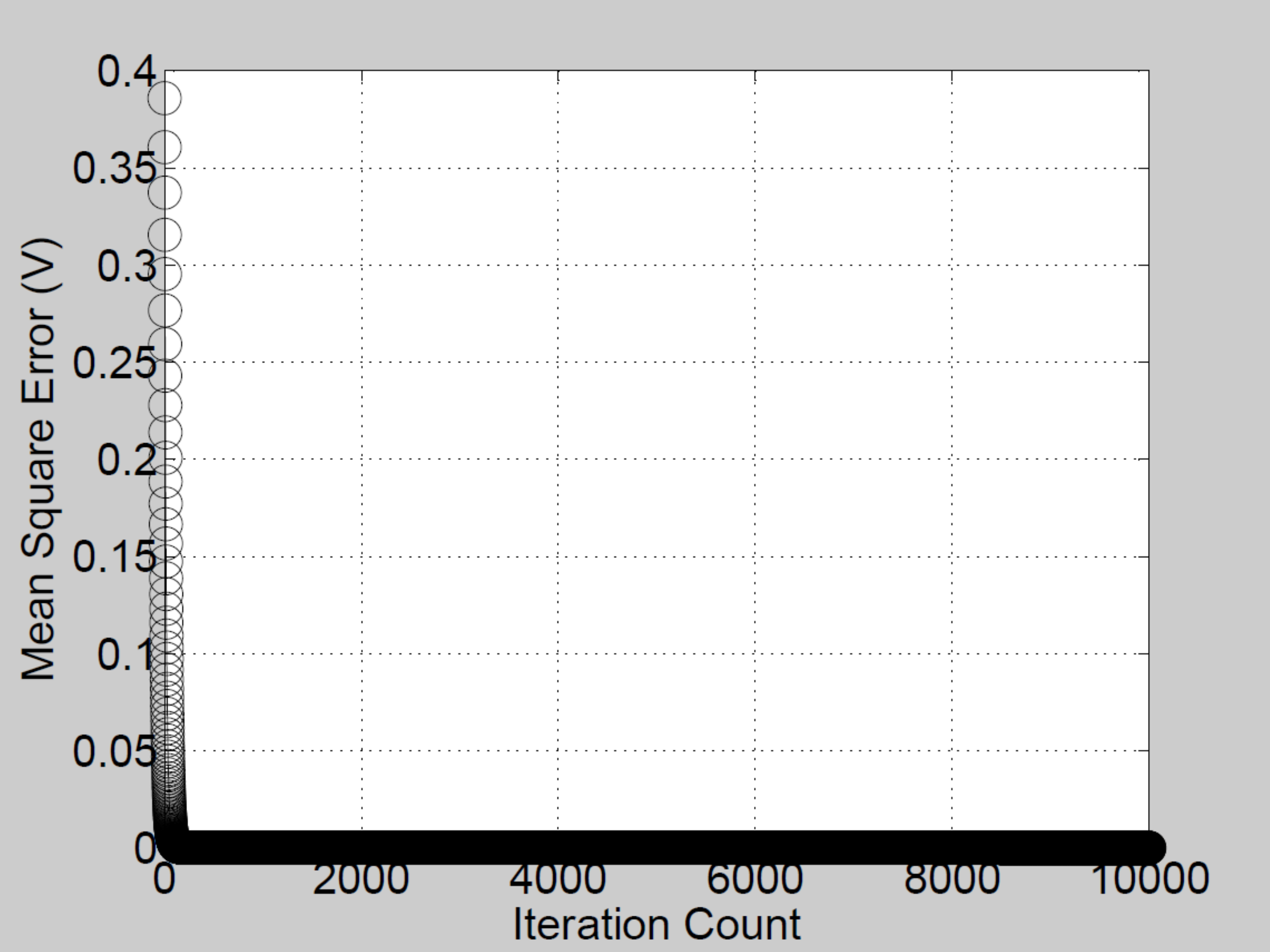} 
& \includegraphics[width=2in]{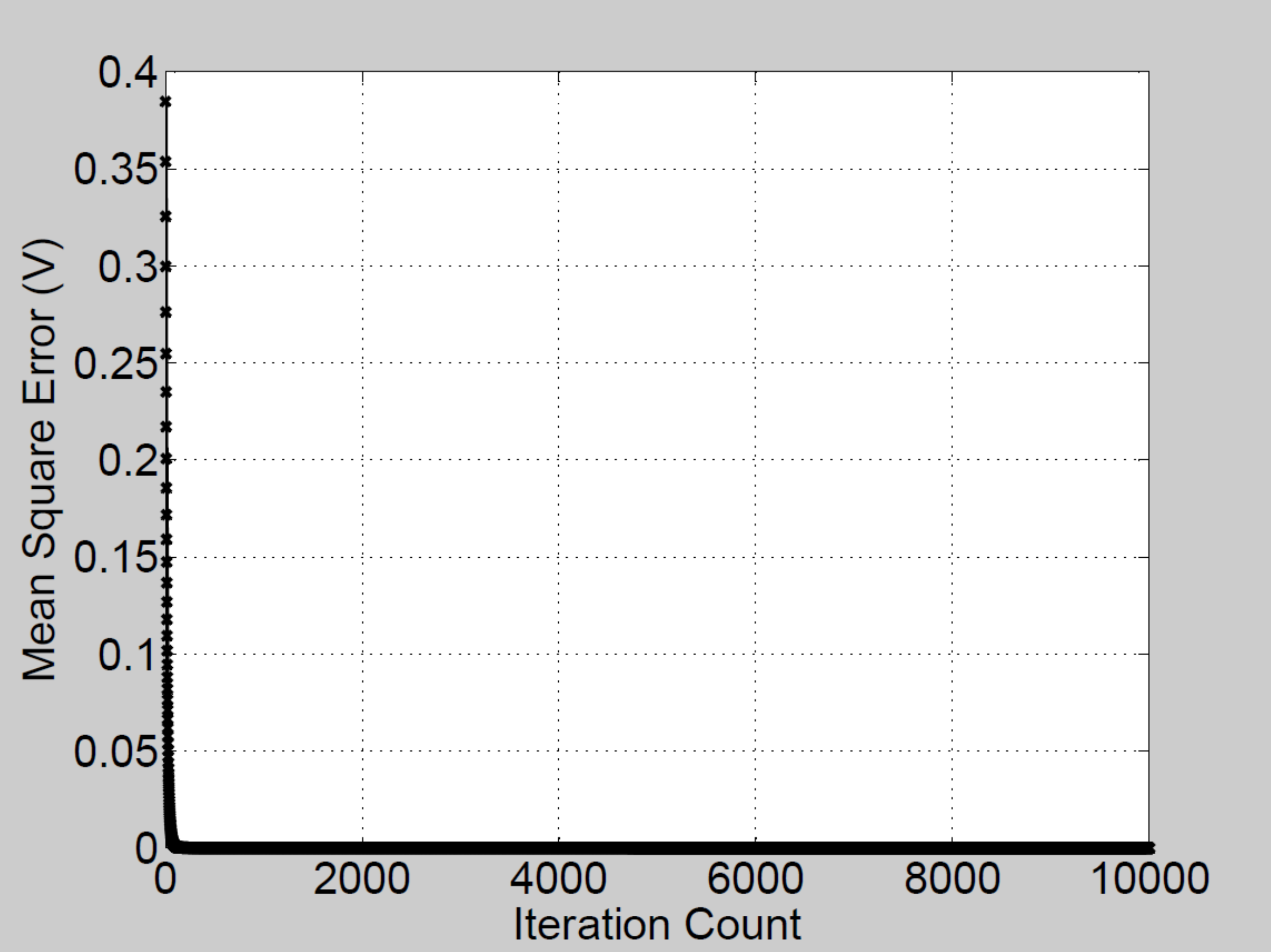} \\
(g): CR: 25. & (h): CR: 51. & (i): CR: 61. \\\\ 
\end{tabular}
\end{center}
\caption{Overall system snapshot for functional expansions (FE) =\{25,51,61\} and their respective convergence rates (CR).} 
\label{fig:sim3}
\end{figure*}

\subsubsection{Stage 2 : Multiplication}

The functionality of the second stage is to perform a 18-bit floating point multiplication of the input signals with their respective weights obtained from the training phase of the FLANN. 
All the expansions originating from the expansion block are fed as input to the multiplication block, which consists of five sub-blocks. 
Out of the five sub-blocks, four of these take 10 inputs and produces 10 outputs, while the fifth block takes 11 inputs and produces 11 outputs.

\subsubsection{Stage 3 : Addition/Subtraction}

The third stage is responsible for performing the addition/subtraction of two 18-bit floating point integers. 
The output from the multiplication blocks are fed as input to this block, which consists of 50 sub-blocks. 
Each sub-block takes as input two 18-bit floating numbers, and produces the added result as output, which is also a 18-bit floating point number.

\section{Evaluation} \label{sec:evaluation}

In this section, we present the results (both of simulation and FPGA implementation) from our FLANN-based nonlinearity compensator model discussed in Sections \ref{sec:system_model} \& \ref{sec:FPGA_impl}.
The experimental data (Table \ref{tab:exp_measured_data}) collected from the LVDT described in Section \ref{sec:pre_exp1} has been utilized for this purpose.

\subsection{Simulation Results} \label{sec:sim}

The simulations have been performed using MATLAB 8.0. 
The comparison metric utilized is the percentage of linear range, which is defined as: 
\begin{equation}	\frac{Number\hspace{1mm}of\hspace{1mm}points\hspace{1mm}in\hspace{1mm}the\hspace{1mm}linear\hspace{1mm}range}{Total\hspace{1mm}number\hspace{1mm}of\hspace{1mm}points}\times 100
\end{equation}
\indent

Figure \ref{fig:sim3}-(a) shows the plot of the input-output characteristics of the LVDT, which is nonlinear. 
Figure \ref{fig:sim3}-(b) shows the inverse input-output characteristic plot from our model simulator. 
Figure \ref{fig:sim3}-(c) shows the final output from the cascade of the LVDT and our model connected in series.
It shows a perfect straight line, which establishes the linearity of the sensor over the entire dynamic range.
\newline
\indent
An important design decision was to choose the number of functional expansion to obtain linearity. 
The value was chosen to be 51, because it was a good trade-off  between less expansions that under-compensate for the nonlinearity (Figure \ref{fig:sim3}-(d)), and more expansions that over-compensate for it (and hence is a waste of resources)(Figure \ref{fig:sim3}-(f)). 
\newline
\indent
Figure \ref{fig:sim3}-(e) shows the overall system response, which demonstrates linear I/O characteristic, after the combination of the FLANN model with the LVDT data. 
Figure \ref{fig:sim3}-(h) shows the convergence rate of our proposed algorithm in terms of mean square error. 
It shows that our algorithm has a fast convergence rate, which converges to its optimal value (neural weights of the FLANN) in its first 100 iterations.
\newline
\indent
Figure \ref{fig:sim3}-(d) shows the overall system response for a FLANN with 25 functional expansions. 
We observe that such a system still shows non-linearity, and the algorithm takes a longer time to converge (Figure \ref{fig:sim3}-(g)).
The response for a system having 61 functional expansion has been shown in Figure \ref{fig:sim3}-(f). 
It shows that this system does not perform better than our system (with 51 expansions), and has comparable convergence rates (Figure \ref{fig:sim3}-(i)). 
\newline
\indent
A comparison among the percentage of linear range achieved with different number of functional expansions has been shown in Table \ref{tab:simulation_studies}.
Thus, in our simulation study, we have achieved $100 \%$ linearity, over the entire range of values collected from our experimental results, in comparison to only $15.38 \%$ from the off-the-shelf LVDT sensor used in our preliminary experiment.

\begin{table}[h]
  \scriptsize
	\centering	
	\caption{Simulation Studies}
	\label{tab:simulation_studies}
		\begin{tabular}{|p{2.25cm}|p{2.25cm}|}
			\hline
		\textbf{Number of functional expansions} & \textbf{\% of linearity range} 
		\\ \hline	\hline
	  11 & 38.46 \\ \hline
	  25 & 84.62 \\ \hline
	  51 & 100.00 \\ \hline
	  61 & 100.00 \\ \hline 	  
	  \end{tabular}
\end{table}

\subsection{FPGA Results} \label{FPGA_results}

\begin{figure}[t]
\begin{center}
\begin{tabular}{c}
\includegraphics[height=2.25in]{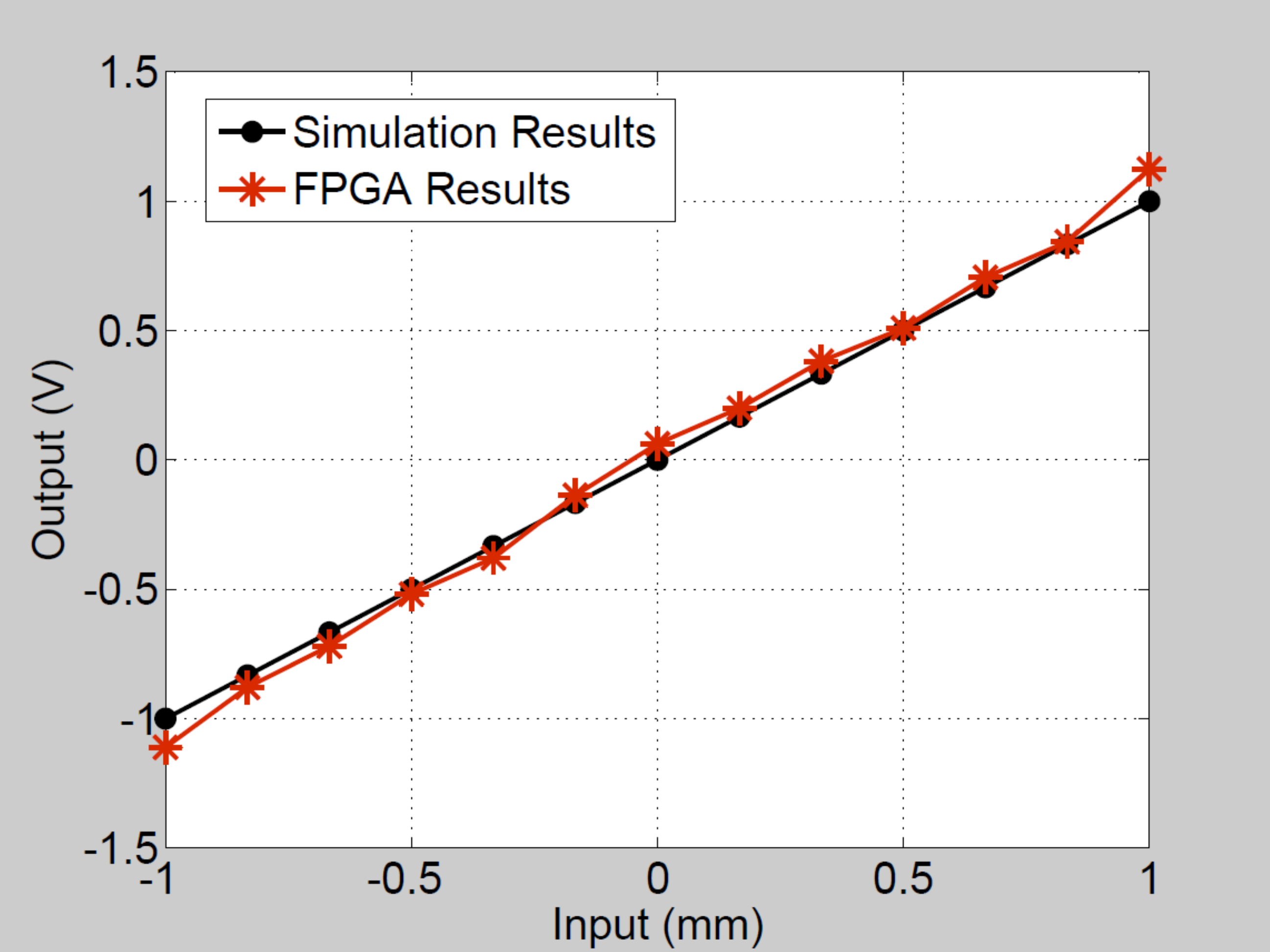} \\
(a)\\
\includegraphics[height=2.25in]{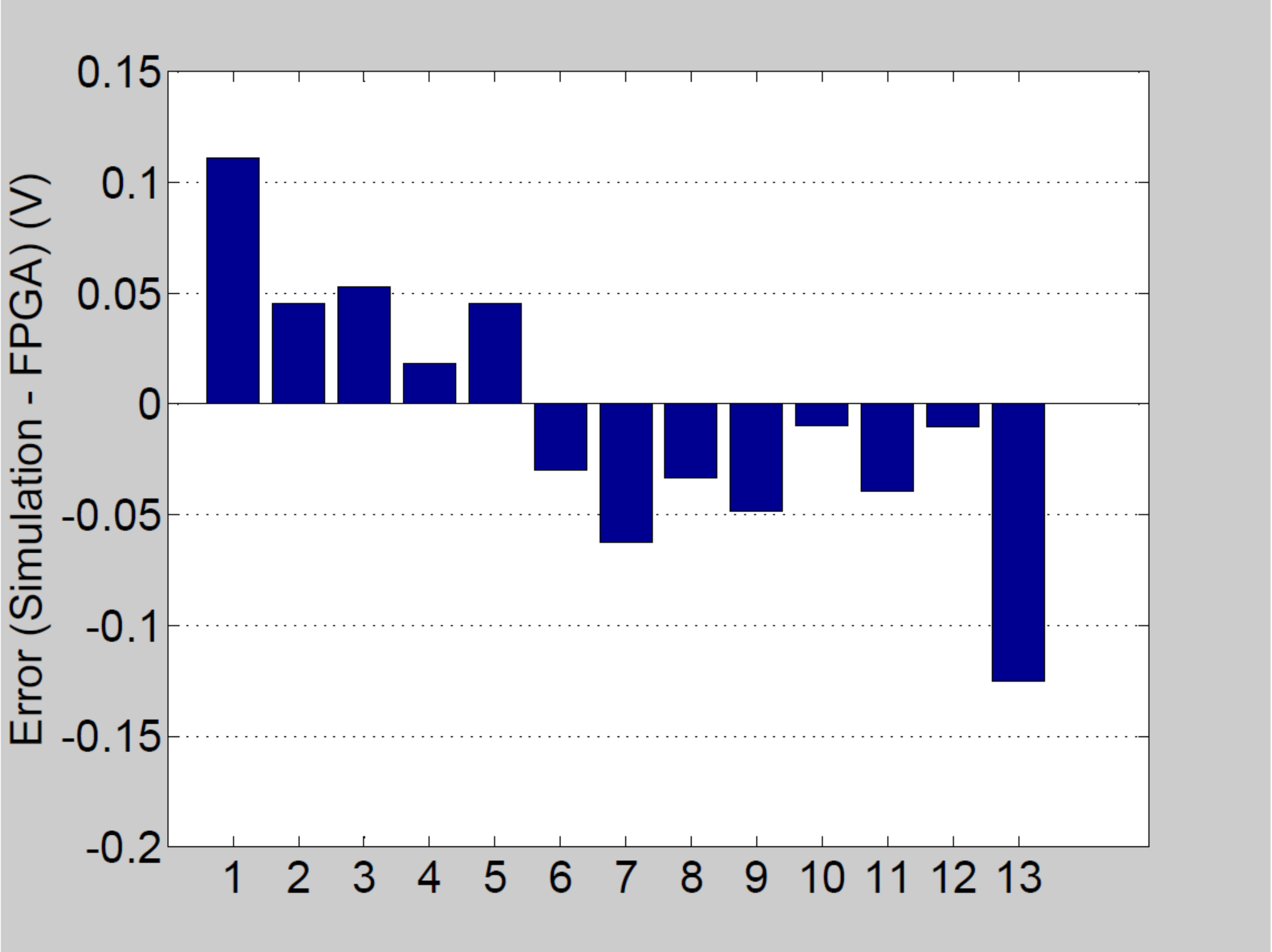}\\ 
(b)\\
\end{tabular}
\end{center}
 \caption{Simulation and FPGA results: (a):Comparison. (b): Error.}
\label{fig:sim4}
\end{figure}

This module was programmed in VHDL \cite{Perry2002} using Xilinx, and subsequently burned on to a SPARTAN-II(PQ208) FPGA. 
The evaluation metric is the error between the perfect (simulation) and real (FPGA) results.
\newline
\indent
The experimentally collected output data from the LVDT (in volts) were provided as input to the FPGA, and its corresponding output values were recorded (which are typically the output from the inverse model). 
The combined values (LVDT (nonlinear) + FPGA (inverse model)) were utilised to generate the final response for the system, which is shown in Figure \ref{fig:sim4}-(a). 
We understand from this figure that the overall input-output characteristics are not perfectly linear, in comparison to the result obtained from the simulation in MATLAB (ideal case). 
Figure \ref{fig:sim4}-(b) gives a plot of the error (i.e. the difference between the simulation and FPGA results), which is mostly below $0.05 V$, except for the end-points. 
\newline 
\indent
This deviation in the FPGA result can be accounted for the fact that minor errors can creep into the system during their hardware implementation, which is not as error free as the simulation environment.
The result obtained from the FPGA implementation is in good agreement with the result that was obtained from the simulation in MATLAB.
Hence, we conclude that the proposed algorithm is feasible, and can be successfully applied for overcoming the non-linearity problem of the LVDT sensor.

\section{Related Work} \label{sec:related_work}

Numerous methods have been reported in literature that have studied the non-linearity problem of an LVDT, and proposed solutions for increasing its linear range. These can be classified as: conventional design techniques, digital signal processing methods and ANN-based inverse models. 
\newline \indent
Saxena et al. in \cite{Saxena1989} utilized the idea of dual secondary coil, due to its insensitivity to the change in excitation frequency and voltage, and proposed a model to verify it. 
The experimental results show that the compensated LVDT provides considerable insensitivity to variation in excitation current, frequency and temperature change, and increases its performance.
However, this technique introduces the tedious effort of dual coil winding that changes the dimension of the coil, and increases the cost and weight of the setup. 
Kano et al. in \cite{Kano1990} used the square coil method to address the nonlinearity of LVDTs.
It proposed the utilization of the perpendicular movement of the core of the axis rather than the conventional parallel movement.
Like the previous method, it also requires a change in the hardware design of the sensor. 
Tian et al. in \cite{Tian1997} proposed algorithms for the design of transducers, and showed that the sensitivity does not depend on the excitation frequency.
However,  these algorithms do not take into account eddy-current effects, and hence can be inaccurate.
\newline 
\indent
Crescini et al. \cite{Crescini1998} proposed a DSP technique based on the spectral estimation of the differential secondary signal for increased accuracy in sensing.
Ford et al. \cite{Ford2001} proposed a DSP-based LVDT signal conditioning system, which had the advantage of improved linearity range, automatic phase correction and better frequency response. 
Flammini et al. in \cite{Flammini2007} proposed a least-mean-square (LMS) based algorithm for fast and accurate position (or displacement) estimations.
The prime advantage of this method is its simplicity in implementation, but it may not be effective to estimate highly dynamic conditions.
Though the DSP techniques yield good results, yet it increases the computation cost of the system, and can only be implemented with the help of dedicated processing boards.
\newline 
\indent
Patra et al. in \cite{Patra2008} has presented the working of ANN-based inverse model to compensate for the nonlinear characteristics in a capacitive pressure sensor.
The adaptive inverse model is used in cascade with the nonlinear sensor to achieve overall linearity.
The standard three-layered multilayer perceptron (MLP) network has been used to develop an adaptive inverse model for these sensors.
However, the MLP-based inverse model involves high computational complexity, and offers unsatisfactory linearity performance.
Mishra et al. in \cite{Mishra2010} proposes a FLANN-based nonlinearity compensator model for LVDT, however, the practicability of such a scheme needs to be verified through a hardware implementation.
\newline
\indent
Our work in this paper proposes a simple FLANN-based inverse model that involves quite less computational complexity than the MLP model. 
Its linearity range is high as compared to MLP. 
Through our FPGA implementation, we have shown that the non-linearity compensator unit can be easily fabricated, unlike a precise coil-winding machine.
A FPGA chip can be designed for realizing this model, and cascaded with the LVDT output for achieving overall linearity.
An initial system design and preliminary work was presented in \cite{Prasant2009:LVDT}.

\section{Future Work and Conclusion} \label{sec:conclusion}

The proposed algorithm utilizes a single FLANN model for the non-linear compensation. 
A possible enhancement can be the application of a cascade of two or more FLANN structures in a single iterative model to achieve better accuracy.
However, it would increase the degree of complexity that may be uncontrollable during its hardware realization. 
Nevertheless, it would be a good analysis to define a trade-off between the number of cascaded FLANNs to their accuracy, complexity and hardware
feasibility.
Presently, the FPGA accepts only a few specific values for generating its inverse characteristics. 
This can be overcome through the use of Systolic architecture design.
It would provide a generic approach to the whole implementation, wherein the system would accept any input value, and process it accordingly.
This scheme, though relatively more complex, will be faster than the Look-up table method.
The overall circuit design needs to been further miniaturized, and checked for power efficiency, for faster response time and durability.
The system efficiency aspects have not been addressed in this paper as it was a proof-of-concept implementation.
\newline
\indent
This paper studies the problem of a LVDT (displacement) sensor.
It presents an experimental study of its non-linearity exhibited in input-output characteristics.
It proposes a FLANN-based inverse modeling approach to compensate for this behavior.
A simulation study of the model was performed in MATLAB.
It proposes an algorithm for the proof-of-concept hardware implementation of this scheme on a SPARTAN-II (PQ208) FPGA using VHDL in Xilinx.
It shows that its implementation is practically feasible on a hardware platform, as the result obtained from the FPGA implementation is in
good agreement with the simulation result in MATLAB.



\bibliographystyle{unsrt}
\bibliography{references} 
%



\end{document}